\documentclass[aps,reprint,superscriptaddress]{revtex4-2}
\usepackage{graphicx,setspace}
\usepackage{dcolumn}
\usepackage{bm,url}
\usepackage[mathlines]{lineno}
\usepackage[linktocpage,colorlinks=true,citecolor=teal,linkcolor=olive,anchorcolor=green,urlcolor=teal]{hyperref}
\usepackage{mathrsfs}
\usepackage{xcolor}
\usepackage{float}
\usepackage[normalem]{ulem}
\usepackage[utf8]{inputenc}
\usepackage{tensor}
\usepackage{physics}
\usepackage{blindtext}
\usepackage{slashed}
\usepackage{tikz}
\usepackage{amsthm,amsfonts,epsfig,natbib,amssymb}
\usepackage{tensor}
\usepackage{mathtools}
\usepackage[scaled]{beramono}
\usepackage[T1]{fontenc}
\usetikzlibrary{decorations.pathmorphing}

\def\be{\begin{equation}}
\def\ee{\end{equation}}
\def\ba{\begin{eqnarray}}
\def\ea{\end{eqnarray}}

\newcommand{\eq}[1]{(\ref{#1})}

\def\q{\theta} \def\w{\omega}  \def\y {\psi}   \def\p {\pi}        \def\k {\kappa} \def\l {\lambda}   \def\c {\chi} \def\b {\beta}    \def\p {\pi}   
        \def\F {\Phi}      \def\.{\cdot}

\begin{document}

\title{Strong Cosmic Censorship in accelerating spacetime}

\author{Ming Zhang}
\email{mingzhang@jxnu.edu.cn}
\affiliation{Department of Physics, Jiangxi Normal University, Nanchang 330022, China}
\author{Jie Jiang}
\email{jiejiang@mail.bnu.edu.cn, corresponding author }
\affiliation{College of Education for the Future, Beijing Normal University, Zhuhai 519087, China}

\begin{abstract}
It has been established that  Strong Cosmic Censorship Conjecture (SCCC)  is respected by an asymptotically flat black hole with charge, but violated by a charged asymptotically de Sitter black hole. We studied the instability of the Cauchy horizon for an accelerating black hole in Einstein theory conformally coupled with a scalar field. The black hole is uncharged while having inner, outer, and acceleration horizons. In the limit of vanishing acceleration, denoted by $A \to 0$, it becomes asymptotically flat and extremal, with its acceleration horizon vanishing as well. By exploring the perturbation of the massless scalar field upon the accelerating black hole, whose decay rate is governed by the quasinormal mode spectra, we show that the SCCC is violated in the near-extremal regime and also in the $A\to 0^+$ limit. Our result is the first observation of a black hole violating the SCCC in an almost asymptotically  Minkowskian flat regime, as well as the first example of a black hole violating the SCCC with neither charge nor rotation.
\end{abstract}

\maketitle



\section{Introduction}
The classical Strong Cosmic Censorship Conjecture  (SCCC), raised by Penrose \cite{Penrose:1969pc}, proposes that the Cauchy horizon of an asymptotically flat black hole in  General Relativity (GR) is not stable and must not form.  This is because initial data on the Cauchy horizon is future inextendible  (see Refs. \cite{rong2022spacetime,Ong:2020xwv} for recent reviews). In other words, the determinism of the GR as a classical field theory logically demands that perturbations of the fields evolve to be divergent on the Cauchy horizon, rendering the Cauchy horizon a mass-inflation singularity. Moreover, the SCCC can be incorporated into a more modern Christodoulou formulation  \cite{Christodoulou:2008nj}, which dictates that even as a weak solution of the Einstein field equation, it is generically impossible to make an extension of a spacetime metric beyond the Cauchy horizon with a locally square-integrable Christoffel symbol.

It has been shown that the asymptotically flat charged Reissner-Nordström (RN) or rotating Kerr black hole respects the SCCC. The reason is that the exponential blueshift effect, which can be determined by the black hole's surface gravity \cite{chandrasekhar1982crossing}, is imposed on the signal sent by an observer outside the black hole  \cite{Simpson:1973ua,Poisson:1990eh,Dafermos:2003wr}. However, it has recently been observed that the RN-de Sitter (RNdS) black hole  violates the SCCC in the near-extremal regime \cite{Cardoso:2017soq,Liu:2019lon}. The reason is that the quasinormal modes (QNMs) of the  massless scalar fields outside the event horizon get a more rapid exponential decay, resulting in the redshift effect suppressing the counterpart blueshift effect. This redshift effect emerges from the existence of a cosmological horizon and is determined by a  spectral gap relating to the imaginary part of the dominant QNMs \cite{Hintz:2016gwb,Hintz:2016jak}. Nonetheless, a cosmological horizon does not necessarily guarantee the violation of the SCCC. The (uncharged and charged) Kerr-de Sitter  black holes, which also own cosmological horizons, were shown to respect the SCCC against scalar and gravitational perturbations \cite{Dias:2018ynt,Hod:2018lmi}. 
	 
Recently, ref. \cite{Destounis:2020yav} pointed out that instead of a cosmological horizon, an acceleration horizon in the charged C metric spacetime \cite{Griffiths:2006tk} (we name it Einstein-Maxwell C metric) can play the same role as a redshift counterbalance. This means that the SCCC may be violated  in the charged accelerating  spacetime in Einstein-Maxwell gravity in the near-extremal regime where the charge-mass ratio approaches 1. In this paper, we will study the stability of the Cauchy horizon for the C metric spacetime in Einstein gravity conformally coupled with a scalar field, which was found in \cite{Charmousis:2009cm} (we name it conformally scalar C metric). 

The Einstein-Maxwell C metric is charged and becomes extremal in a large electric charge regime. However, the conformally scalar C metric we will study is uncharged, and intriguingly, it becomes extremal in the regime of vanishing acceleration. As a result, the former C metric is not asymptotically Minkowskian flat in the extremal regime due to the existence of the acceleration horizon. Conversely, the latter one is asymptotically Minkowskian flat in the extremal regime (and in other words, it is {\it{almost}} asymptotically Minkowskian flat in the near-extremal regime). In fact, no good reason can be appointed to choose either conformal coupling or minimal coupling of the scalar field to the spacetime curvature in nature \cite{carroll2019spacetime}. However, the conformal coupling is more interesting from two points of view \cite{Winstanley:2002jt}: first, there exist non-trivial solutions in asymptotically flat spacetime for this type of coupling; second, in the investigation of  quantum field theory in the curved spacetime, the conformal coupling is the most natural selection. 

What we want to emphasize is that all previous  investigations have shown that asymptotically flat black holes respect the SCCC. However, we will show that the conformally scalar C metric, which represents a conformally scalar accelerating black hole, can violate the SCCC in the near-extremal regime, where we can have the acceleration of the black hole $A\to 0^+$. This almost vanishing acceleration, in turn, makes the black hole almost asymptotically Minkowskian flat without an acceleration horizon. 



\section{Conformally  scalar accelerating black hole}
We consider the Einstein theory conformally coupled with a scalar field $\Psi$,
\begin{equation}
I=\frac{1}{16\pi}\int \mathrm{d}^{4} x \sqrt{-g}\left[R-\kappa\left(\nabla_{\mu} \Psi \nabla^{\mu} \Psi+\frac{R}{6} \Psi^{2}\right)
\right],
\end{equation}
where $\kappa=8\pi$, $R$ is the Ricci scalar, and we have used geometric  units $c=G=1$.    A conformally scalar accelerating black hole solution of the above equations of motion is \cite{Charmousis:2009cm}
\begin{equation}
\begin{aligned}\label{met}
\bar{g}_{\mu\nu}=& \frac{1}{\Omega(r, \theta)^{2}}\left[-f(r) d t^{2}+\frac{d r^{2}}{f(r)}+\frac{r^{2}}{P(\theta)} d \theta^{2}\right.\\
&\left.+r^{2} \sin ^{2} \theta P(\theta) d \phi^{2}\right],
\end{aligned}
\end{equation}
\begin{equation}
\Omega(r, \theta)=1+A r \cos\theta,
\end{equation}
\begin{equation}\label{black}
f(r)=\left(\frac{1}{r^2}-A^{2}\right)(r-M)\left(r-\frac{M}{1+2 A M}\right),
\end{equation}
\begin{equation}
P(\theta)=(1+A M \cos \theta)\left(1+\frac{A M}{1+2 A M} \cos \theta\right),
\end{equation}
\begin{equation}\label{Psieq}
\Psi(r, \theta)=\sqrt{\frac{6}{\kappa}} \frac{M(A r \cos \theta+1)}{r(1+A M)+M(A r \cos \theta-1)},
\end{equation}
where $M$ is the mass parameter, $A$ stands for the acceleration of the black hole.  Intriguingly, the scalar field $\Psi$ only plays an indirect role in the metric.

The Cauchy horizon, event horizon, and acceleration horizon of the black hole are given by \cite{Charmousis:2009cm}
\begin{equation}
r_{-}=\frac{M}{1+2 A M}, \quad r_{+}=M, \quad r_{A}=\frac{1}{A}.
\end{equation}
We denote the surface gravity at each horizon  as
\begin{equation}
\kappa_{i} \equiv \frac{\left|f^{\prime}\left(r_{i}\right)\right|}{2},
\end{equation}
where $i\in\left\{-, +, A\right\}$. To maitain the causality of the spacetime, these horizons should satisfy $0\leq r_{-} \leq r_{+} \leq r_{A}$. Additionally, to ensure the right signature, we must ensure that $P(\theta)>0$ for $\theta \in[0, \pi]$. There are conical deficits on the poles where $\theta=0,\,\pi$, as the periodicity of the coordinate $\phi\neq 2\pi$. Setting $\phi \in[0,2 \pi K)$ there, with $K$ being a dimensionless constant, yields the deficit angles
\begin{equation}\label{ptm}
\varpi_{\pm}=2 \pi\left[1-K P_{\theta}^{\pm}\right],
\end{equation}
where $P_\theta^+=P(0)$ and $P_\theta^-=P(\pi)$. One of  the deficit angles, say $\varpi_{+}$, can be removed by making $K=1 / P_{\theta}^{+}$. The deficit angles can be viewed as cosmic strings connecting the black hole with the asymptotic infinity \cite{Emparan:1999wa}.



\section{Scalar perturbation of the accelerating black hole}
We now consider a massless neutral scalar field $\bar{\varphi}$ perturbating upon the accelerating black hole (\ref{met}), which can be depicted by the Klein-Gordon wave equation
\begin{equation}
\square_{\bar{g}} \bar{\varphi}=0,
\end{equation}
where we have denoted $\square_{\bar{g}}\equiv\bar{g}_{\mu \nu} \bar{\nabla}^{\mu} \bar{\nabla}^{\nu}$ with $\bar{\nabla}_\mu$ being the covariant derivative associated with the metric (\ref{met}). As the scalar curvature $\bar{R}$ of the  spacetime vanishes, though the spacetime is not asymptotically Minkowskian due to the existence of the acceleration horizon, the above wave equation can thus be equivalently transformed into a conformal-transformation-invariant form
\begin{equation}
\left(\square_{\bar{g}}-\frac{1}{6}\bar{R}\right) \bar{\varphi}=0.
\end{equation}
	
To make this equation separable for the components of the scalar field function $\bar{\varphi}$, we take a conformal gauge \cite{Hawking:1997ia,wald2010general}
\begin{equation}
g_{\mu \nu} \rightarrow \Omega^{2} \bar{g}_{\mu \nu},\quad \varphi \rightarrow \Omega^{-1} \bar{\varphi}.
\end{equation}
By doing so, the equation of motion for the scalar field reads
\begin{equation}\label{ckgeq}
\left(\square_{g}-\frac{1}{6}R\right) \varphi=0,
\end{equation}
where $\square_{g}\equiv g_{\mu \nu} \nabla^{\mu} \nabla^{\nu}$ with $\nabla_\mu$ being the covariant derivative associated with the conformal metric $g_{\mu \nu}$, and $R$ is the Ricci scalar of the conformal metric $g_{\mu\nu}$.

We perform a field decomposition of  $\varphi$ as
\begin{equation}\label{vphif}
\varphi=e^{-i \omega t} e^{i m \phi} \frac{\Phi(r)}{r} \chi(\theta),
\end{equation}
where $\omega\equiv\omega_R+i \omega_I $ represents the frequency of the mode, and its imaginary part determines whether the field is convergent or divergent. Considering the periodicity of the coordinate $\phi$, we obtain the azimuthal harmonic index  $m=m_0 P_\theta^-$ [cf. eq. (\ref{ptm})], where $m_0$ is an integer. Without loss of generality, we can assume that $m_0\geq 0$.	

Consequently, we can decompose eq. (\ref{ckgeq})  into two separate equations:
\begin{equation}\label{ei1}
\frac{d^{2} \Phi(r)}{d r_{*}^{2}}+\left(\omega^{2}-V_{r}\right) \Phi(r)=0,
\end{equation}
\begin{equation}\label{ei2}
\frac{d^{2} \chi(\theta)}{d z^{2}}-\left(m^{2}-V_{\theta}\right) \chi(\theta)=0.
\end{equation}
 We have taken transformations of the coordinates by $d r_{*}\equiv d r / f(r)$ and $d z\equiv d \theta /(P(\theta) \sin \theta)$. The radial and angular effective potentials respectively are
\begin{equation}
    V_r=f(r)\left(\frac{\lambda}{r^2}-\frac{f(r)}{3 r^2}+\frac{f^{\prime}(r)}{3 r}-\frac{f^{\prime \prime}(r)}{6}\right),
\end{equation}
\begin{equation}
    \begin{aligned}
    V_\theta= & P(\theta)\left(\lambda \sin ^2 \theta-\frac{P(\theta) \sin ^2 \theta}{3}\right. \\
    & \left.+\frac{\sin \theta \cos \theta P^{\prime}(\theta)}{2}+\frac{\sin ^2 \theta P^{\prime \prime}(\theta)}{6}\right),
    \end{aligned}
\end{equation}
where $\lambda$ is a separation constant between the radial and angular equations. The symbol ${}^{\prime}$ denotes derivatives with respect to the variables $r$ and $\theta$ of the functions $f(r)$ and $P(\theta)$, respectively.

To solve eqs. (\ref{ei1}) and (\ref{ei2}), additional physical boundary conditions should be imposed, namely,
\ba\begin{aligned}\label{abc1}
\Phi(r) \sim \begin{cases}e^{-i \omega r_{*}}, & r_{*} \rightarrow-\infty\,\,\left(r \rightarrow r_{+}\right), \\ e^{+i \omega r_{*}}, & r_{*} \rightarrow+\infty\,\,\left(r \rightarrow r_{A}\right),\end{cases}
\end{aligned}\ea
and
\begin{equation}\label{abc2}
\chi(\theta) \sim \begin{cases}e^{+m z}, & z \rightarrow-\infty\,\,(\theta \rightarrow 0), \\ e^{-m z}, & z \rightarrow+\infty\,\,(\theta \rightarrow \pi)\end{cases}
\end{equation}
eq. (\ref{abc1}) means that the QNMs are purely ingoing crossing the event horizon and purely outgoing crossing the acceleration horizon. eq. (\ref{abc2}) ensures that the wave solution is finite at the $\theta$-boundary. For a given $m_0$, the separation constant $\lambda$ can be obtained by solving eqs. (\ref{ei2}) and (\ref{abc2}), then the QNMs $\omega$ can be calculated by substituting $\lambda$ into eqs. (\ref{ei1}) and (\ref{abc1}). Furthermore, it has been shown in ref. \cite{Kofron:2015gli} that in the limit $A\to 0$, due to the conformal coupling, we have $\lambda=l(l+1)+1/3$, which relates the eigenvalue $\lambda$ with the spherical harmonic index $l$. Therefore, we can map each $\lambda$ of the accelerating black hole obtained by solving eq. \eq{ei2} to certain harmonic indexes $m_0$ and $l$.



\section{Strong cosmic censorship conjecture}
 To determine whether or not the SCCC is respected by the conformally scalar accelerating black hole, we should consider a control parameter
\begin{equation}
\beta \equiv \alpha / \kappa_{-},
\end{equation}
where \cite{Maeda:1999sv,Dafermos:2012np}
\begin{equation}\label{info}
    \alpha \equiv-\text{inf} _{m n}\{\operatorname{Im}(\omega)\}
    \end{equation}
with $n$ being the overtone number. As has been proven in ref. \cite{Destounis:2020yav}, the critical control condition that ensures the first-order derivative of $\varphi$ is locally square integrable at $r\to r_-$ should be $\beta>1 / 2$. Thus, we will now first numerically calculate the featured QNMs of the accelerating black hole against the scalar field perturbation, which is governed by eq. (\ref{ckgeq}), and then observe how the control parameter behaves.

To numerically obtain the separation constant and the QNM frequencies \cite{Konoplya:2011qq}, we use the direct integration method based on the \textit{NDSolve} function of \textit{Wolfram@ Mathematica} \cite{Chandrasekhar:1975zza,Molina:2010fb}. We verify our results based on the pseudospectral method \cite{Jansen:2017oag,Miguel:2020uln}.

\begin{figure*}[htbp!]
    \centering
    \includegraphics[width=0.4\textwidth]{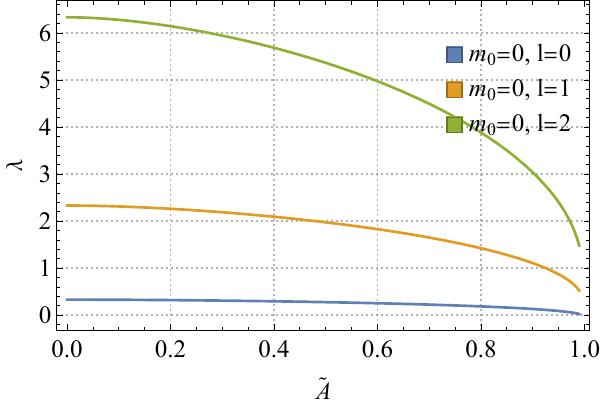}   
    \includegraphics[width=0.41\textwidth]{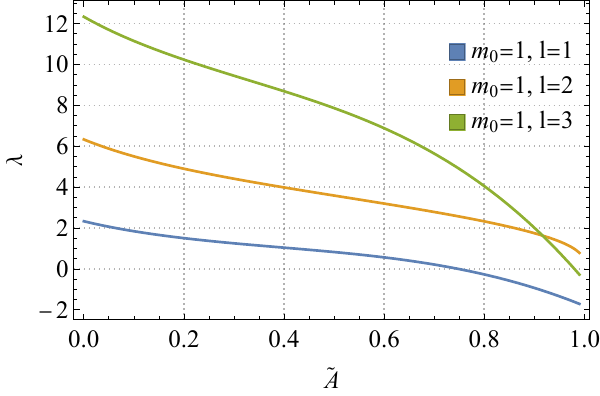}
    \caption{Variations of the separation constant $\lambda$ with respect to the reduced acceleration $\tilde{A}$ for the accelerating black hole.}
    \label{fig1}
\end{figure*}

\begin{table*}[ht]
\caption{ Comparison between the direct integration method and pseudospectral method used for the calculation of the control parameter  $\beta=-{\omega_I}/{\kappa_-}$ with various harmonic indexes $m_0$ and $l$ for the accelerating black hole with $\tilde{A}=0.2$.}\label{tab1}
    \centering
    \renewcommand\arraystretch{1.8}
\begin{tabular}{ccccc}
\hline Numerical method & $\beta (m_0=0, l=0)$ & $\beta (m_0=0, l=1)$ & $\beta (m_0=1, l=1)$ & $\beta (m_0=10, l=10)$ \\
 \hline
    Direct integration method& $0.381561761293738$ & $0.374026754942024$ & $0.37472180750644$ & $0.37270437679659$ \\

    Pseudospectral method & $0.381561761289558$ & $0.374026754942638$ & $0.37472180750529$ & $0.37270446932970$\\
\hline
\end{tabular}
\end{table*}

\begin{figure*}
    \centering
    \includegraphics[width=0.4\textwidth]{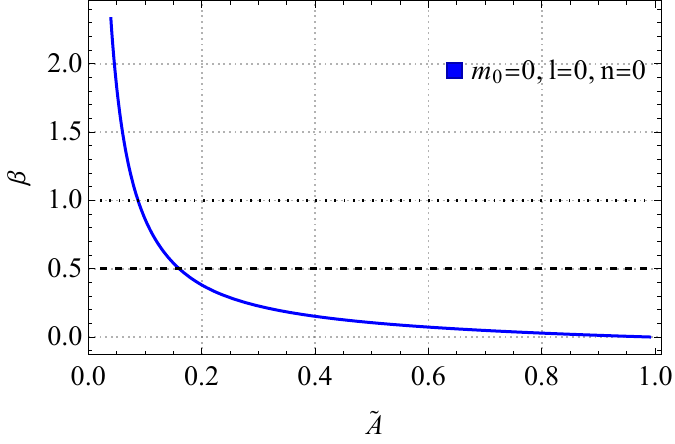}
        \includegraphics[width=0.4\textwidth]{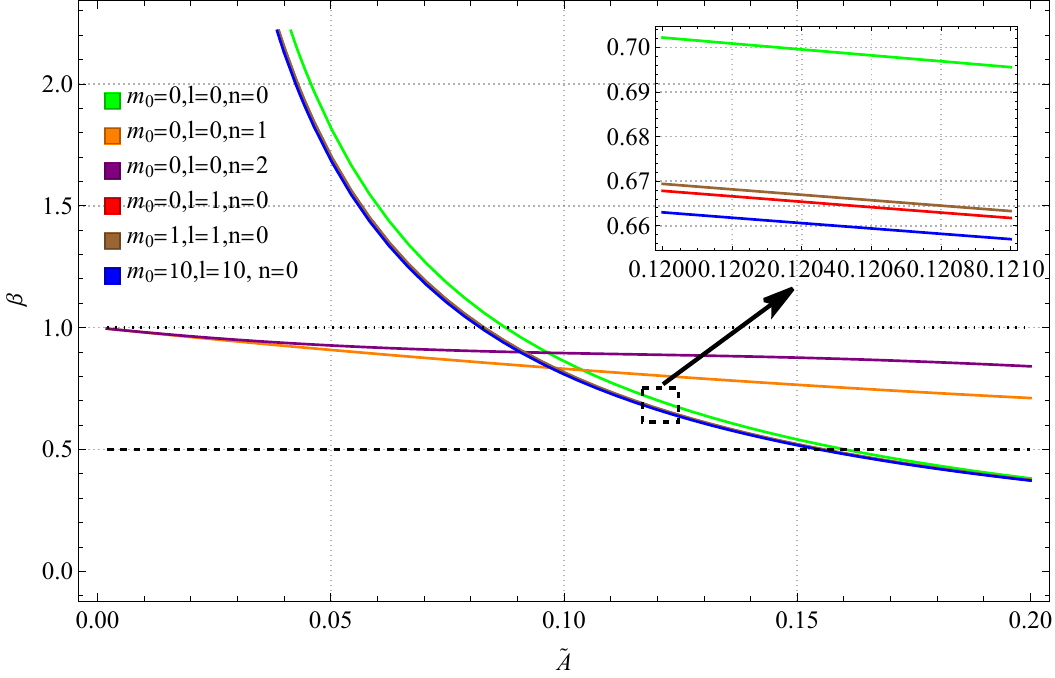}
    \caption{The left diagram shows the value of $\b=-\w_I/\k_{-}$ for the dominate acceleration modes with $m_0=0, l=0$ as a function of the reduced acceleration $\tilde{A}$. The dashed and dotted horizontal lines denote $\b=1/2$ and $\b=1$, respectively. In the right diagram, the blue cure represents the dominant modes with $(m_0=10, l=10)$, which approximates to the photon sphere mode with large $\l$, the orange curve corresponds to $(m_0=0, l=0)$ dominant near-extremal modes, and the green, red, and brown curves correspond to the first $(m_0=0, l=0)$, $(m_0=0, l=1)$ and $(m_0=1, l=1)$ modes, respectively. }
    \label{fig2}
\end{figure*}

First of all, we calculate the separation constant $\l$ in eq. \eq{ei2} with the boundary condition \eq{abc2}. Note that $\chi(\theta)$ is singular at $\q=0, \p$. To eliminate these singularities, we rescale $\c(\q)$ by
\ba\begin{aligned}\label{cct}
\c(\q)=\left[\sin\frac{\q}{2}\right]^{m/P(0)}\left[\cos\frac{\q}{2}\right]^{m/P(\p)}\tilde{\c}(\q)\,.
\end{aligned}\ea
The equation for $\tilde{\c}(\q)$ can be obtained by substituting eq. \eq{cct} into eq. \eq{ei2}. We do not show the explicit expression of this equation as it is too lengthy. From the boundary condition \eq{abc2}, we see that the physical solution demands that $\tilde{\c}(\q)$ be regular at $\q=0$ and $\q=\p$. That is to say, we can expand $\tilde{\c}(\q)$ as
\begin{equation}
    \tilde{\chi}(\theta)=\sum_{n=0}^{\infty} b_n \theta^n, \quad \theta \rightarrow 0,
\end{equation}
\begin{equation}
        \tilde{\chi}(\theta)=\sum_{n=0}^{\infty} d_n(\theta-\pi)^n, \quad \theta \rightarrow \pi
\end{equation}
near $\q=0$ and $\q=\p$ with $b_n$ and $d_n$ the expansion coefficients. Substituting these coefficients into the rescaled equation, we can obtain the expansion coefficients and we can easily see that the higher-order coefficients are determined by the zeroth-order coefficients $b_0$ and $d_0$. Then, we can use the series solutions to evaluate values of $\tilde{\c}(\q)$ and $d\tilde{\c}/d\theta$  near $\q=0, \p$, and use them as initial values to solve the rescaled  equation of $\tilde{\chi}(\theta)$. Using the initial value near $\q=0$, we can solve the rescaled equation in the interval $(0, \p/2)$ based on the \textit{NDSolve} function of \textit{Wolfram@ Mathematica}. Similarly, we can also solve the rescaled equation in the interval $(\p/2, \p)$ using the initial value near $\q=\p$. Then, the acceptable solution of $\tilde{\c}(\q)$ and the separation constant $\l$ are obtained by requiring that $\tilde{\c}(\q)$ and its first-order derivative function are continuous at $\q=\p/2$.

We show the separation constant $\l$ as a function of the reduced dimensionless acceleration $\tilde{A}\equiv A M$, with different harmonic indexes $m_0$ and $l$, in Fig. \ref{fig1}. From this figure, we can see that the separation constant $\l$ monotonically decreases as $\tilde{A}$ increases for a given set of harmonic indexes. This means that the separation constant $\lambda$ is unique for a certain mode and spacetime parameter.

Next, we will evaluate the QNM frequencies using the direct integration method. Similar to our consideration of  $\c(\q)$, we also rescale the radial function $\F(r)$ as
\be\begin{aligned}\label{respsi}
\F(r)=\left(\frac{r-r_+}{r_+}\right)^{\frac{i \omega}{2\kappa_+}}\left(\frac{r_A-r}{r_A}\right)^{-\frac{i\omega}{2\kappa_A}}\psi(x),
\end{aligned}\ee
where
\be
r=r_++(r_A-r_+)x\,.
\ee
Then, the rescaled field $\y(x)$ is regular at its domain $x\in [0,1]$ and we can easily obtain its series solution near the event horizon $(x=0)$ and the acceleration horizon $(x=1)$. Based on this, we can perform the direct integration method to obtain the QNMs. 

To ensure the reliability, in Table \ref{tab1} we compare the results of the ratio $\b=\alpha / \kappa_{-}=-\w_I/\k_-$ [cf. eq. (\ref{info})] evaluated by the direct integration method with those obtained by the pseudospectral method for various values of  harmonic indexes $m_0$ and $l$, for the accelerating black hole with $\tilde{A}=0.2$. The results show  consistency between the two methods. In the left diagram of Fig. \ref{fig2}, we show the value of $\b$ as a function of the reduced acceleration $\tilde{A}$ for $m_0=0,\,l=0,\,n=0$. To identify the dominant modes, we also consider the overtone numbers  $n=0, \,1,\, 2$ in the right diagram. From this figure, it is evident that for large $\tilde{A}$, we always have $\b<1/2$, indicating that  the SCCC is valid in this region. Hence, we only need to examine the slowly accelerating regime to determine the validity of the SCCC. In the right diagram of Fig. \ref{fig2}, we display the representative families of modes that might provide the dominant QNMs, i.e., the first three branches of $(m_0=0, l=0)$ modes,  the first branch of  $(m_0=0, l=1)$ mode, $(m_0=1, l=1)$ mode and $(m_0=10, l=10)$ mode. From the figure, we can see that the dominant modes comprise the second branch of the $(m_0=0, l=0)$ mode with $n=1$, and the first branch of the $(m_0=10, l=10)$ mode with $n=0$. The second branch of the $(m_0=0, l=0)$ mode corresponds to the near-extremal mode that yields  purely imaginary frequencies and governs the ringdown for an approaching extremal black hole, i.e., in the $\tilde{A}\to 0^+$ limit. The first branch of the $(m_0=10, l=10)$ mode approximates to the photon sphere mode that corresponds to the QNMs with large $l$. As demonstrated in Fig. \ref{fig2}, we can find that for a slowly accelerating black hole, the dominant mode is determined by the near-extremal mode; however, with sufficiently large acceleration, the dominant mode is given by the photon sphere mode. In summary, we can see that there exist some parametric regions of the accelerating black hole in which their dominant modes make the control parameter $\b>1/2$, implying that the SCCC is violated. 



\section{Closing remarks}
Supermassive accelerating black holes could reside in the center of galaxies \cite{morris2017nonthermal,Vilenkin:2018zol} and their velocities  must be small $(\lesssim 100 \mathrm{~km} / \mathrm{s})$ \cite{Ashoorioon:2021znw}. In this paper, we have analyzed the issue of  the Cauchy horizon instability for the accelerating black hole in the Einstein theory conformally coupled with a scalar field. The black hole and the conformal scalar are only characterized by two parameters -- mass and acceleration. However, it has three horizons: the Cauchy horizon, the event horizon, and the acceleration horizon.  When the acceleration of the black hole decreases to zero, the former two horizons merge, resulting in an extremal black hole. Intriguingly, at the same time, the last horizon extends to spatial infinity, rendering the black hole asymptotically Minkowskian flat.

We have tested Christodoulou's formulation of SCCC in this spacetime and found it to be violated in the near-extremal regime, or in other words, in the slow acceleration regime. To this end, we studied the scalar perturbation of the accelerating black hole and then calculated the QNMs of this process. The meaning of our result can be summarized as follows. On the one hand, all previous investigations have indicated that SCCC is respected for asymptotically flat initial data which is close to the RN or the Kerr. In our case, when the acceleration is zero, the black hole transforms into an extremal RN-like spacetime [however, it is not precisely an extremal RN as it still has conformally coupled scalar hair, as shown in eqs. (\ref{black}) and (\ref{Psieq})] or a BPS (and extremal) spacetime in the supergravity theory. It should be noted that string theory achieved great success in the realm of quantum gravity through its ability to calculate the microscopic entropy for a class of supersymmetric, asymptotically flat extremal RN-like black holes including multiple scalars within the supergravity theories \cite{Strominger:1996sh,Hristov:2022pmo,Cai:2001jc}. Our outcome suggests that the SCCC can be violated in a black hole spacetime that is almost asymptotically Minkowskian flat in the near-extremal regime with a vanishing acceleration ($A\to 0^+$).

On the other hand, to the best of our knowledge, our result is the first one that violates Christodoulou's formulation of SCCC with a black hole that takes neither charge nor rotation. Mass is the only traditional conserved quantity that act as an hair in the black hole solution. The dimensional acceleration parameter, though must be incorporated into the thermodynamic first law of the black hole \cite{Appels:2016uha,Ashoorioon:2021znw}, does not relate to the Killing vector of the spacetime. 

Finally, we would like to comment that: first, though Kerr-Newman family black holes are widely accepted toy models of GR, it is valuable to study other modified gravity models that provide viable alternatives against the predictions of GR \cite{Cvetic:2019ekr,Chong:2005hr,Cvetic:2022mal} and the Einstein theory conformally coupled with a scalar field is special, as we stated in the Introduction section; secondly, it is worthwhile to extend the present analysis to the stationary case, which is more astrophysically relevant.

\begin{acknowledgments}
 M. Z. is supported by the National Natural Science Foundation of China with Grant No. 12005080.  J. J.  is supported by the National Natural Science Foundation of China with Grant No. 210510101, the Guangdong Basic and Applied Research Foundation with Grant No. 217200003, and the Talents Introduction Foundation of Beijing Normal University with Grant No. 310432102.
\end{acknowledgments}

\bibliography{refs.bib}

\begin{thebibliography}{40}%
\makeatletter
\providecommand \@ifxundefined [1]{%
 \@ifx{#1\undefined}
}%
\providecommand \@ifnum [1]{%
 \ifnum #1\expandafter \@firstoftwo
 \else \expandafter \@secondoftwo
 \fi
}%
\providecommand \@ifx [1]{%
 \ifx #1\expandafter \@firstoftwo
 \else \expandafter \@secondoftwo
 \fi
}%
\providecommand \natexlab [1]{#1}%
\providecommand \enquote  [1]{``#1''}%
\providecommand \bibnamefont  [1]{#1}%
\providecommand \bibfnamefont [1]{#1}%
\providecommand \citenamefont [1]{#1}%
\providecommand \href@noop [0]{\@secondoftwo}%
\providecommand \href [0]{\begingroup \@sanitize@url \@href}%
\providecommand \@href[1]{\@@startlink{#1}\@@href}%
\providecommand \@@href[1]{\endgroup#1\@@endlink}%
\providecommand \@sanitize@url [0]{\catcode `\\12\catcode `\$12\catcode
  `\&12\catcode `\#12\catcode `\^12\catcode `\_12\catcode `\%12\relax}%
\providecommand \@@startlink[1]{}%
\providecommand \@@endlink[0]{}%
\providecommand \url  [0]{\begingroup\@sanitize@url \@url }%
\providecommand \@url [1]{\endgroup\@href {#1}{\urlprefix }}%
\providecommand \urlprefix  [0]{URL }%
\providecommand \Eprint [0]{\href }%
\providecommand \doibase [0]{https://doi.org/}%
\providecommand \selectlanguage [0]{\@gobble}%
\providecommand \bibinfo  [0]{\@secondoftwo}%
\providecommand \bibfield  [0]{\@secondoftwo}%
\providecommand \translation [1]{[#1]}%
\providecommand \BibitemOpen [0]{}%
\providecommand \bibitemStop [0]{}%
\providecommand \bibitemNoStop [0]{.\EOS\space}%
\providecommand \EOS [0]{\spacefactor3000\relax}%
\providecommand \BibitemShut  [1]{\csname bibitem#1\endcsname}%
\let\auto@bib@innerbib\@empty
\bibitem [{\citenamefont {Penrose}(1969)}]{Penrose:1969pc}%
  \BibitemOpen
  \bibfield  {author} {\bibinfo {author} {\bibfnamefont {R.}~\bibnamefont
  {Penrose}},\ }\href@noop {} {\bibfield  {journal} {\bibinfo  {journal} {Riv.
  Nuovo Cim.}\ }\textbf {\bibinfo {volume} {1}},\ \bibinfo {pages} {252}
  (\bibinfo {year} {1969})}\BibitemShut {NoStop}%
\bibitem [{\citenamefont {Cai}\ \emph {et~al.}(2022)\citenamefont {Cai},
  \citenamefont {Cao}, \citenamefont {Li},\ and\ \citenamefont
  {Yang}}]{rong2022spacetime}%
  \BibitemOpen
  \bibfield  {author} {\bibinfo {author} {\bibfnamefont {R.-G.}\ \bibnamefont
  {Cai}}, \bibinfo {author} {\bibfnamefont {L.-M.}\ \bibnamefont {Cao}},
  \bibinfo {author} {\bibfnamefont {L.}~\bibnamefont {Li}},\ and\ \bibinfo
  {author} {\bibfnamefont {R.-Q.}\ \bibnamefont {Yang}},\ }\href
  {https://doi.org/10.1360/SSPMA-2022-0069} {\bibfield  {journal} {\bibinfo
  {journal} {Sci. Sin.-Phys. Mech. Astron.}\ }\textbf {\bibinfo {volume}
  {52}},\ \bibinfo {pages} {110401} (\bibinfo {year} {2022})}\BibitemShut
  {NoStop}%
\bibitem [{\citenamefont {Ong}(2020)}]{Ong:2020xwv}%
  \BibitemOpen
  \bibfield  {author} {\bibinfo {author} {\bibfnamefont {Y.~C.}\ \bibnamefont
  {Ong}},\ }\href {https://doi.org/10.1142/S0217751X20300070} {\bibfield
  {journal} {\bibinfo  {journal} {Int. J. Mod. Phys. A}\ }\textbf {\bibinfo
  {volume} {35}},\ \bibinfo {pages} {14} (\bibinfo {year} {2020})},\ \Eprint
  {https://arxiv.org/abs/2005.07032} {arXiv:2005.07032 [gr-qc]} \BibitemShut
  {NoStop}%
\bibitem [{\citenamefont {Christodoulou}(2008)}]{Christodoulou:2008nj}%
  \BibitemOpen
  \bibfield  {author} {\bibinfo {author} {\bibfnamefont {D.}~\bibnamefont
  {Christodoulou}},\ }in\ \href {https://doi.org/10.1142/9789814374552_0002}
  {\emph {\bibinfo {booktitle} {{12th Marcel Grossmann Meeting on General
  Relativity}}}}\ (\bibinfo {year} {2008})\ \Eprint
  {https://arxiv.org/abs/0805.3880} {arXiv:0805.3880 [gr-qc]} \BibitemShut
  {NoStop}%
\bibitem [{\citenamefont {Chandrasekhar}\ and\ \citenamefont
  {Hartle}(1982)}]{chandrasekhar1982crossing}%
  \BibitemOpen
  \bibfield  {author} {\bibinfo {author} {\bibfnamefont {S.}~\bibnamefont
  {Chandrasekhar}}\ and\ \bibinfo {author} {\bibfnamefont {J.~B.}\ \bibnamefont
  {Hartle}},\ }\href {https://doi.org/10.1098/rspa.1982.0160} {\bibfield
  {journal} {\bibinfo  {journal} {Proceedings of the Royal Society of London.
  A. Mathematical and Physical Sciences}\ }\textbf {\bibinfo {volume} {384}},\
  \bibinfo {pages} {301} (\bibinfo {year} {1982})}\BibitemShut {NoStop}%
\bibitem [{\citenamefont {Simpson}\ and\ \citenamefont
  {Penrose}(1973)}]{Simpson:1973ua}%
  \BibitemOpen
  \bibfield  {author} {\bibinfo {author} {\bibfnamefont {M.}~\bibnamefont
  {Simpson}}\ and\ \bibinfo {author} {\bibfnamefont {R.}~\bibnamefont
  {Penrose}},\ }\href {https://doi.org/10.1007/BF00792069} {\bibfield
  {journal} {\bibinfo  {journal} {Int. J. Theor. Phys.}\ }\textbf {\bibinfo
  {volume} {7}},\ \bibinfo {pages} {183} (\bibinfo {year} {1973})}\BibitemShut
  {NoStop}%
\bibitem [{\citenamefont {Poisson}\ and\ \citenamefont
  {Israel}(1990)}]{Poisson:1990eh}%
  \BibitemOpen
  \bibfield  {author} {\bibinfo {author} {\bibfnamefont {E.}~\bibnamefont
  {Poisson}}\ and\ \bibinfo {author} {\bibfnamefont {W.}~\bibnamefont
  {Israel}},\ }\href {https://doi.org/10.1103/PhysRevD.41.1796} {\bibfield
  {journal} {\bibinfo  {journal} {Phys. Rev. D}\ }\textbf {\bibinfo {volume}
  {41}},\ \bibinfo {pages} {1796} (\bibinfo {year} {1990})}\BibitemShut
  {NoStop}%
\bibitem [{\citenamefont {Dafermos}(2005)}]{Dafermos:2003wr}%
  \BibitemOpen
  \bibfield  {author} {\bibinfo {author} {\bibfnamefont {M.}~\bibnamefont
  {Dafermos}},\ }\href {https://doi.org/10.1002/cpa.20071} {\bibfield
  {journal} {\bibinfo  {journal} {Commun. Pure Appl. Math.}\ }\textbf {\bibinfo
  {volume} {58}},\ \bibinfo {pages} {0445} (\bibinfo {year} {2005})},\ \Eprint
  {https://arxiv.org/abs/gr-qc/0307013} {arXiv:gr-qc/0307013} \BibitemShut
  {NoStop}%
\bibitem [{\citenamefont {Cardoso}\ \emph {et~al.}(2018)\citenamefont
  {Cardoso}, \citenamefont {Costa}, \citenamefont {Destounis}, \citenamefont
  {Hintz},\ and\ \citenamefont {Jansen}}]{Cardoso:2017soq}%
  \BibitemOpen
  \bibfield  {author} {\bibinfo {author} {\bibfnamefont {V.}~\bibnamefont
  {Cardoso}}, \bibinfo {author} {\bibfnamefont {J.~a.~L.}\ \bibnamefont
  {Costa}}, \bibinfo {author} {\bibfnamefont {K.}~\bibnamefont {Destounis}},
  \bibinfo {author} {\bibfnamefont {P.}~\bibnamefont {Hintz}},\ and\ \bibinfo
  {author} {\bibfnamefont {A.}~\bibnamefont {Jansen}},\ }\href
  {https://doi.org/10.1103/PhysRevLett.120.031103} {\bibfield  {journal}
  {\bibinfo  {journal} {Phys. Rev. Lett.}\ }\textbf {\bibinfo {volume} {120}},\
  \bibinfo {pages} {031103} (\bibinfo {year} {2018})},\ \Eprint
  {https://arxiv.org/abs/1711.10502} {arXiv:1711.10502 [gr-qc]} \BibitemShut
  {NoStop}%
\bibitem [{\citenamefont {Liu}\ \emph {et~al.}(2019)\citenamefont {Liu},
  \citenamefont {Tang}, \citenamefont {Destounis}, \citenamefont {Wang},
  \citenamefont {Papantonopoulos},\ and\ \citenamefont {Zhang}}]{Liu:2019lon}%
  \BibitemOpen
  \bibfield  {author} {\bibinfo {author} {\bibfnamefont {H.}~\bibnamefont
  {Liu}}, \bibinfo {author} {\bibfnamefont {Z.}~\bibnamefont {Tang}}, \bibinfo
  {author} {\bibfnamefont {K.}~\bibnamefont {Destounis}}, \bibinfo {author}
  {\bibfnamefont {B.}~\bibnamefont {Wang}}, \bibinfo {author} {\bibfnamefont
  {E.}~\bibnamefont {Papantonopoulos}},\ and\ \bibinfo {author} {\bibfnamefont
  {H.}~\bibnamefont {Zhang}},\ }\href {https://doi.org/10.1007/JHEP03(2019)187}
  {\bibfield  {journal} {\bibinfo  {journal} {JHEP}\ }\textbf {\bibinfo
  {volume} {03}},\ \bibinfo {pages} {187}},\ \Eprint
  {https://arxiv.org/abs/1902.01865} {arXiv:1902.01865 [gr-qc]} \BibitemShut
  {NoStop}%
\bibitem [{\citenamefont {Hintz}\ and\ \citenamefont
  {Vasy}(2016)}]{Hintz:2016gwb}%
  \BibitemOpen
  \bibfield  {author} {\bibinfo {author} {\bibfnamefont {P.}~\bibnamefont
  {Hintz}}\ and\ \bibinfo {author} {\bibfnamefont {A.}~\bibnamefont {Vasy}}\
  }\href {https://doi.org/10.4310/acta.2018.v220.n1.a1}
  {10.4310/acta.2018.v220.n1.a1} (\bibinfo {year} {2016}),\ \Eprint
  {https://arxiv.org/abs/1606.04014} {arXiv:1606.04014 [math.DG]} \BibitemShut
  {NoStop}%
\bibitem [{\citenamefont {Hintz}(2016)}]{Hintz:2016jak}%
  \BibitemOpen
  \bibfield  {author} {\bibinfo {author} {\bibfnamefont {P.}~\bibnamefont
  {Hintz}}\ }\href {https://doi.org/10.1007/s40818-018-0047-y}
  {10.1007/s40818-018-0047-y} (\bibinfo {year} {2016}),\ \Eprint
  {https://arxiv.org/abs/1612.04489} {arXiv:1612.04489 [math.AP]} \BibitemShut
  {NoStop}%
\bibitem [{\citenamefont {Dias}\ \emph {et~al.}(2018)\citenamefont {Dias},
  \citenamefont {Eperon}, \citenamefont {Reall},\ and\ \citenamefont
  {Santos}}]{Dias:2018ynt}%
  \BibitemOpen
  \bibfield  {author} {\bibinfo {author} {\bibfnamefont {O.~J.~C.}\
  \bibnamefont {Dias}}, \bibinfo {author} {\bibfnamefont {F.~C.}\ \bibnamefont
  {Eperon}}, \bibinfo {author} {\bibfnamefont {H.~S.}\ \bibnamefont {Reall}},\
  and\ \bibinfo {author} {\bibfnamefont {J.~E.}\ \bibnamefont {Santos}},\
  }\href {https://doi.org/10.1103/PhysRevD.97.104060} {\bibfield  {journal}
  {\bibinfo  {journal} {Phys. Rev. D}\ }\textbf {\bibinfo {volume} {97}},\
  \bibinfo {pages} {104060} (\bibinfo {year} {2018})},\ \Eprint
  {https://arxiv.org/abs/1801.09694} {arXiv:1801.09694 [gr-qc]} \BibitemShut
  {NoStop}%
\bibitem [{\citenamefont {Hod}(2018)}]{Hod:2018lmi}%
  \BibitemOpen
  \bibfield  {author} {\bibinfo {author} {\bibfnamefont {S.}~\bibnamefont
  {Hod}},\ }\href {https://doi.org/10.1016/j.physletb.2018.03.020} {\bibfield
  {journal} {\bibinfo  {journal} {Phys. Lett. B}\ }\textbf {\bibinfo {volume}
  {780}},\ \bibinfo {pages} {221} (\bibinfo {year} {2018})},\ \Eprint
  {https://arxiv.org/abs/1803.05443} {arXiv:1803.05443 [gr-qc]} \BibitemShut
  {NoStop}%
\bibitem [{\citenamefont {Destounis}\ \emph {et~al.}(2020)\citenamefont
  {Destounis}, \citenamefont {Fontana},\ and\ \citenamefont
  {Mena}}]{Destounis:2020yav}%
  \BibitemOpen
  \bibfield  {author} {\bibinfo {author} {\bibfnamefont {K.}~\bibnamefont
  {Destounis}}, \bibinfo {author} {\bibfnamefont {R.~D.~B.}\ \bibnamefont
  {Fontana}},\ and\ \bibinfo {author} {\bibfnamefont {F.~C.}\ \bibnamefont
  {Mena}},\ }\href {https://doi.org/10.1103/PhysRevD.102.104037} {\bibfield
  {journal} {\bibinfo  {journal} {Phys. Rev. D}\ }\textbf {\bibinfo {volume}
  {102}},\ \bibinfo {pages} {104037} (\bibinfo {year} {2020})},\ \Eprint
  {https://arxiv.org/abs/2006.01152} {arXiv:2006.01152 [gr-qc]} \BibitemShut
  {NoStop}%
\bibitem [{\citenamefont {Griffiths}\ \emph {et~al.}(2006)\citenamefont
  {Griffiths}, \citenamefont {Krtous},\ and\ \citenamefont
  {Podolsky}}]{Griffiths:2006tk}%
  \BibitemOpen
  \bibfield  {author} {\bibinfo {author} {\bibfnamefont {J.~B.}\ \bibnamefont
  {Griffiths}}, \bibinfo {author} {\bibfnamefont {P.}~\bibnamefont {Krtous}},\
  and\ \bibinfo {author} {\bibfnamefont {J.}~\bibnamefont {Podolsky}},\ }\href
  {https://doi.org/10.1088/0264-9381/23/23/008} {\bibfield  {journal} {\bibinfo
   {journal} {Class. Quant. Grav.}\ }\textbf {\bibinfo {volume} {23}},\
  \bibinfo {pages} {6745} (\bibinfo {year} {2006})},\ \Eprint
  {https://arxiv.org/abs/gr-qc/0609056} {arXiv:gr-qc/0609056} \BibitemShut
  {NoStop}%
\bibitem [{\citenamefont {Charmousis}\ \emph {et~al.}(2009)\citenamefont
  {Charmousis}, \citenamefont {Kolyvaris},\ and\ \citenamefont
  {Papantonopoulos}}]{Charmousis:2009cm}%
  \BibitemOpen
  \bibfield  {author} {\bibinfo {author} {\bibfnamefont {C.}~\bibnamefont
  {Charmousis}}, \bibinfo {author} {\bibfnamefont {T.}~\bibnamefont
  {Kolyvaris}},\ and\ \bibinfo {author} {\bibfnamefont {E.}~\bibnamefont
  {Papantonopoulos}},\ }\href {https://doi.org/10.1088/0264-9381/26/17/175012}
  {\bibfield  {journal} {\bibinfo  {journal} {Class. Quant. Grav.}\ }\textbf
  {\bibinfo {volume} {26}},\ \bibinfo {pages} {175012} (\bibinfo {year}
  {2009})},\ \Eprint {https://arxiv.org/abs/0906.5568} {arXiv:0906.5568
  [gr-qc]} \BibitemShut {NoStop}%
\bibitem [{\citenamefont {Carroll}(2019)}]{carroll2019spacetime}%
  \BibitemOpen
  \bibfield  {author} {\bibinfo {author} {\bibfnamefont {S.~M.}\ \bibnamefont
  {Carroll}},\ }\href@noop {} {\emph {\bibinfo {title} {Spacetime and
  geometry}}}\ (\bibinfo  {publisher} {Cambridge University Press},\ \bibinfo
  {year} {2019})\BibitemShut {NoStop}%
\bibitem [{\citenamefont {Winstanley}(2003)}]{Winstanley:2002jt}%
  \BibitemOpen
  \bibfield  {author} {\bibinfo {author} {\bibfnamefont {E.}~\bibnamefont
  {Winstanley}},\ }\href {https://doi.org/10.1023/A:1022871809835} {\bibfield
  {journal} {\bibinfo  {journal} {Found. Phys.}\ }\textbf {\bibinfo {volume}
  {33}},\ \bibinfo {pages} {111} (\bibinfo {year} {2003})},\ \Eprint
  {https://arxiv.org/abs/gr-qc/0205092} {arXiv:gr-qc/0205092} \BibitemShut
  {NoStop}%
\bibitem [{\citenamefont {Emparan}\ \emph {et~al.}(2000)\citenamefont
  {Emparan}, \citenamefont {Horowitz},\ and\ \citenamefont
  {Myers}}]{Emparan:1999wa}%
  \BibitemOpen
  \bibfield  {author} {\bibinfo {author} {\bibfnamefont {R.}~\bibnamefont
  {Emparan}}, \bibinfo {author} {\bibfnamefont {G.~T.}\ \bibnamefont
  {Horowitz}},\ and\ \bibinfo {author} {\bibfnamefont {R.~C.}\ \bibnamefont
  {Myers}},\ }\href {https://doi.org/10.1088/1126-6708/2000/01/007} {\bibfield
  {journal} {\bibinfo  {journal} {JHEP}\ }\textbf {\bibinfo {volume} {01}},\
  \bibinfo {pages} {007}},\ \Eprint {https://arxiv.org/abs/hep-th/9911043}
  {arXiv:hep-th/9911043} \BibitemShut {NoStop}%
\bibitem [{\citenamefont {Hawking}\ and\ \citenamefont
  {Ross}(1997)}]{Hawking:1997ia}%
  \BibitemOpen
  \bibfield  {author} {\bibinfo {author} {\bibfnamefont {S.~W.}\ \bibnamefont
  {Hawking}}\ and\ \bibinfo {author} {\bibfnamefont {S.~F.}\ \bibnamefont
  {Ross}},\ }\href {https://doi.org/10.1103/PhysRevD.56.6403} {\bibfield
  {journal} {\bibinfo  {journal} {Phys. Rev. D}\ }\textbf {\bibinfo {volume}
  {56}},\ \bibinfo {pages} {6403} (\bibinfo {year} {1997})},\ \Eprint
  {https://arxiv.org/abs/hep-th/9705147} {arXiv:hep-th/9705147} \BibitemShut
  {NoStop}%
\bibitem [{\citenamefont {Wald}(2010)}]{wald2010general}%
  \BibitemOpen
  \bibfield  {author} {\bibinfo {author} {\bibfnamefont {R.~M.}\ \bibnamefont
  {Wald}},\ }\href@noop {} {\emph {\bibinfo {title} {General relativity}}}\
  (\bibinfo  {publisher} {University of Chicago press},\ \bibinfo {year}
  {2010})\BibitemShut {NoStop}%
\bibitem [{\citenamefont {Kofrov{n}}(2015)}]{Kofron:2015gli}%
  \BibitemOpen
  \bibfield  {author} {\bibinfo {author} {\bibfnamefont {D.}~\bibnamefont
  {Kofrov{n}}},\ }\href {https://doi.org/10.1103/PhysRevD.92.124064} {\bibfield
   {journal} {\bibinfo  {journal} {Phys. Rev. D}\ }\textbf {\bibinfo {volume}
  {92}},\ \bibinfo {pages} {124064} (\bibinfo {year} {2015})},\ \Eprint
  {https://arxiv.org/abs/1603.01451} {arXiv:1603.01451 [gr-qc]} \BibitemShut
  {NoStop}%
\bibitem [{\citenamefont {Maeda}\ \emph {et~al.}(2000)\citenamefont {Maeda},
  \citenamefont {Torii},\ and\ \citenamefont {Narita}}]{Maeda:1999sv}%
  \BibitemOpen
  \bibfield  {author} {\bibinfo {author} {\bibfnamefont {K.}~\bibnamefont
  {Maeda}}, \bibinfo {author} {\bibfnamefont {T.}~\bibnamefont {Torii}},\ and\
  \bibinfo {author} {\bibfnamefont {M.}~\bibnamefont {Narita}},\ }\href
  {https://doi.org/10.1103/PhysRevD.61.024020} {\bibfield  {journal} {\bibinfo
  {journal} {Phys. Rev. D}\ }\textbf {\bibinfo {volume} {61}},\ \bibinfo
  {pages} {024020} (\bibinfo {year} {2000})},\ \Eprint
  {https://arxiv.org/abs/gr-qc/9908007} {arXiv:gr-qc/9908007} \BibitemShut
  {NoStop}%
\bibitem [{\citenamefont {Dafermos}(2014)}]{Dafermos:2012np}%
  \BibitemOpen
  \bibfield  {author} {\bibinfo {author} {\bibfnamefont {M.}~\bibnamefont
  {Dafermos}},\ }\href {https://doi.org/10.1007/s00220-014-2063-4} {\bibfield
  {journal} {\bibinfo  {journal} {Commun. Math. Phys.}\ }\textbf {\bibinfo
  {volume} {332}},\ \bibinfo {pages} {729} (\bibinfo {year} {2014})},\ \Eprint
  {https://arxiv.org/abs/1201.1797} {arXiv:1201.1797 [gr-qc]} \BibitemShut
  {NoStop}%
\bibitem [{\citenamefont {Konoplya}\ and\ \citenamefont
  {Zhidenko}(2011)}]{Konoplya:2011qq}%
  \BibitemOpen
  \bibfield  {author} {\bibinfo {author} {\bibfnamefont {R.~A.}\ \bibnamefont
  {Konoplya}}\ and\ \bibinfo {author} {\bibfnamefont {A.}~\bibnamefont
  {Zhidenko}},\ }\href {https://doi.org/10.1103/RevModPhys.83.793} {\bibfield
  {journal} {\bibinfo  {journal} {Rev. Mod. Phys.}\ }\textbf {\bibinfo {volume}
  {83}},\ \bibinfo {pages} {793} (\bibinfo {year} {2011})},\ \Eprint
  {https://arxiv.org/abs/1102.4014} {arXiv:1102.4014 [gr-qc]} \BibitemShut
  {NoStop}%
\bibitem [{\citenamefont {Chandrasekhar}\ and\ \citenamefont
  {Detweiler}(1975)}]{Chandrasekhar:1975zza}%
  \BibitemOpen
  \bibfield  {author} {\bibinfo {author} {\bibfnamefont {S.}~\bibnamefont
  {Chandrasekhar}}\ and\ \bibinfo {author} {\bibfnamefont {S.~L.}\ \bibnamefont
  {Detweiler}},\ }\href {https://doi.org/10.1098/rspa.1975.0112} {\bibfield
  {journal} {\bibinfo  {journal} {Proc. Roy. Soc. Lond. A}\ }\textbf {\bibinfo
  {volume} {344}},\ \bibinfo {pages} {441} (\bibinfo {year}
  {1975})}\BibitemShut {NoStop}%
\bibitem [{\citenamefont {Molina}\ \emph {et~al.}(2010)\citenamefont {Molina},
  \citenamefont {Pani}, \citenamefont {Cardoso},\ and\ \citenamefont
  {Gualtieri}}]{Molina:2010fb}%
  \BibitemOpen
  \bibfield  {author} {\bibinfo {author} {\bibfnamefont {C.}~\bibnamefont
  {Molina}}, \bibinfo {author} {\bibfnamefont {P.}~\bibnamefont {Pani}},
  \bibinfo {author} {\bibfnamefont {V.}~\bibnamefont {Cardoso}},\ and\ \bibinfo
  {author} {\bibfnamefont {L.}~\bibnamefont {Gualtieri}},\ }\href
  {https://doi.org/10.1103/PhysRevD.81.124021} {\bibfield  {journal} {\bibinfo
  {journal} {Phys. Rev. D}\ }\textbf {\bibinfo {volume} {81}},\ \bibinfo
  {pages} {124021} (\bibinfo {year} {2010})},\ \Eprint
  {https://arxiv.org/abs/1004.4007} {arXiv:1004.4007 [gr-qc]} \BibitemShut
  {NoStop}%
\bibitem [{\citenamefont {Jansen}(2017)}]{Jansen:2017oag}%
  \BibitemOpen
  \bibfield  {author} {\bibinfo {author} {\bibfnamefont {A.}~\bibnamefont
  {Jansen}},\ }\href {https://doi.org/10.1140/epjp/i2017-11825-9} {\bibfield
  {journal} {\bibinfo  {journal} {Eur. Phys. J. Plus}\ }\textbf {\bibinfo
  {volume} {132}},\ \bibinfo {pages} {546} (\bibinfo {year} {2017})},\ \Eprint
  {https://arxiv.org/abs/1709.09178} {arXiv:1709.09178 [gr-qc]} \BibitemShut
  {NoStop}%
\bibitem [{\citenamefont {Miguel}(2021)}]{Miguel:2020uln}%
  \BibitemOpen
  \bibfield  {author} {\bibinfo {author} {\bibfnamefont {F.~S.}\ \bibnamefont
  {Miguel}},\ }\href {https://doi.org/10.1103/PhysRevD.103.064077} {\bibfield
  {journal} {\bibinfo  {journal} {Phys. Rev. D}\ }\textbf {\bibinfo {volume}
  {103}},\ \bibinfo {pages} {064077} (\bibinfo {year} {2021})},\ \Eprint
  {https://arxiv.org/abs/2012.10455} {arXiv:2012.10455 [gr-qc]} \BibitemShut
  {NoStop}%
\bibitem [{\citenamefont {Morris}\ \emph {et~al.}(2017)\citenamefont {Morris},
  \citenamefont {Zhao},\ and\ \citenamefont {Goss}}]{morris2017nonthermal}%
  \BibitemOpen
  \bibfield  {author} {\bibinfo {author} {\bibfnamefont {M.~R.}\ \bibnamefont
  {Morris}}, \bibinfo {author} {\bibfnamefont {J.-H.}\ \bibnamefont {Zhao}},\
  and\ \bibinfo {author} {\bibfnamefont {W.}~\bibnamefont {Goss}},\ }\href
  {https://doi.org/10.3847/2041-8213/aa9985} {\bibfield  {journal} {\bibinfo
  {journal} {The Astrophysical Journal Letters}\ }\textbf {\bibinfo {volume}
  {850}},\ \bibinfo {pages} {L23} (\bibinfo {year} {2017})}\BibitemShut
  {NoStop}%
\bibitem [{\citenamefont {Vilenkin}\ \emph {et~al.}(2018)\citenamefont
  {Vilenkin}, \citenamefont {Levin},\ and\ \citenamefont
  {Gruzinov}}]{Vilenkin:2018zol}%
  \BibitemOpen
  \bibfield  {author} {\bibinfo {author} {\bibfnamefont {A.}~\bibnamefont
  {Vilenkin}}, \bibinfo {author} {\bibfnamefont {Y.}~\bibnamefont {Levin}},\
  and\ \bibinfo {author} {\bibfnamefont {A.}~\bibnamefont {Gruzinov}},\ }\href
  {https://doi.org/10.1088/1475-7516/2018/11/008} {\bibfield  {journal}
  {\bibinfo  {journal} {JCAP}\ }\textbf {\bibinfo {volume} {11}},\ \bibinfo
  {pages} {008}},\ \Eprint {https://arxiv.org/abs/1808.00670} {arXiv:1808.00670
  [astro-ph.CO]} \BibitemShut {NoStop}%
\bibitem [{\citenamefont {Ashoorioon}\ \emph {et~al.}(2022)\citenamefont
  {Ashoorioon}, \citenamefont {Jahani~Poshteh},\ and\ \citenamefont
  {Mann}}]{Ashoorioon:2021znw}%
  \BibitemOpen
  \bibfield  {author} {\bibinfo {author} {\bibfnamefont {A.}~\bibnamefont
  {Ashoorioon}}, \bibinfo {author} {\bibfnamefont {M.~B.}\ \bibnamefont
  {Jahani~Poshteh}},\ and\ \bibinfo {author} {\bibfnamefont {R.~B.}\
  \bibnamefont {Mann}},\ }\href
  {https://doi.org/10.1103/PhysRevLett.129.031102} {\bibfield  {journal}
  {\bibinfo  {journal} {Phys. Rev. Lett.}\ }\textbf {\bibinfo {volume} {129}},\
  \bibinfo {pages} {031102} (\bibinfo {year} {2022})},\ \Eprint
  {https://arxiv.org/abs/2110.13132} {arXiv:2110.13132 [gr-qc]} \BibitemShut
  {NoStop}%
\bibitem [{\citenamefont {Strominger}\ and\ \citenamefont
  {Vafa}(1996)}]{Strominger:1996sh}%
  \BibitemOpen
  \bibfield  {author} {\bibinfo {author} {\bibfnamefont {A.}~\bibnamefont
  {Strominger}}\ and\ \bibinfo {author} {\bibfnamefont {C.}~\bibnamefont
  {Vafa}},\ }\href {https://doi.org/10.1016/0370-2693(96)00345-0} {\bibfield
  {journal} {\bibinfo  {journal} {Phys. Lett. B}\ }\textbf {\bibinfo {volume}
  {379}},\ \bibinfo {pages} {99} (\bibinfo {year} {1996})},\ \Eprint
  {https://arxiv.org/abs/hep-th/9601029} {arXiv:hep-th/9601029} \BibitemShut
  {NoStop}%
\bibitem [{\citenamefont {Hristov}(2022)}]{Hristov:2022pmo}%
  \BibitemOpen
  \bibfield  {author} {\bibinfo {author} {\bibfnamefont {K.}~\bibnamefont
  {Hristov}},\ }\href {https://doi.org/10.1007/JHEP09(2022)204} {\bibfield
  {journal} {\bibinfo  {journal} {JHEP}\ }\textbf {\bibinfo {volume} {09}},\
  \bibinfo {pages} {204}},\ \Eprint {https://arxiv.org/abs/2207.12437}
  {arXiv:2207.12437 [hep-th]} \BibitemShut {NoStop}%
\bibitem [{\citenamefont {Cai}(2001)}]{Cai:2001jc}%
  \BibitemOpen
  \bibfield  {author} {\bibinfo {author} {\bibfnamefont {R.-G.}\ \bibnamefont
  {Cai}},\ }\href {https://doi.org/10.1103/PhysRevD.63.124018} {\bibfield
  {journal} {\bibinfo  {journal} {Phys. Rev. D}\ }\textbf {\bibinfo {volume}
  {63}},\ \bibinfo {pages} {124018} (\bibinfo {year} {2001})},\ \Eprint
  {https://arxiv.org/abs/hep-th/0102113} {arXiv:hep-th/0102113} \BibitemShut
  {NoStop}%
\bibitem [{\citenamefont {Appels}\ \emph {et~al.}(2016)\citenamefont {Appels},
  \citenamefont {Gregory},\ and\ \citenamefont {Kubiznak}}]{Appels:2016uha}%
  \BibitemOpen
  \bibfield  {author} {\bibinfo {author} {\bibfnamefont {M.}~\bibnamefont
  {Appels}}, \bibinfo {author} {\bibfnamefont {R.}~\bibnamefont {Gregory}},\
  and\ \bibinfo {author} {\bibfnamefont {D.}~\bibnamefont {Kubiznak}},\ }\href
  {https://doi.org/10.1103/PhysRevLett.117.131303} {\bibfield  {journal}
  {\bibinfo  {journal} {Phys. Rev. Lett.}\ }\textbf {\bibinfo {volume} {117}},\
  \bibinfo {pages} {131303} (\bibinfo {year} {2016})},\ \Eprint
  {https://arxiv.org/abs/1604.08812} {arXiv:1604.08812 [hep-th]} \BibitemShut
  {NoStop}%
\bibitem [{\citenamefont {Cvetic}\ \emph {et~al.}(2020)\citenamefont {Cvetic},
  \citenamefont {Gibbons}, \citenamefont {Pope},\ and\ \citenamefont
  {Whiting}}]{Cvetic:2019ekr}%
  \BibitemOpen
  \bibfield  {author} {\bibinfo {author} {\bibfnamefont {M.}~\bibnamefont
  {Cvetic}}, \bibinfo {author} {\bibfnamefont {G.~W.}\ \bibnamefont {Gibbons}},
  \bibinfo {author} {\bibfnamefont {C.~N.}\ \bibnamefont {Pope}},\ and\
  \bibinfo {author} {\bibfnamefont {B.~F.}\ \bibnamefont {Whiting}},\ }\href
  {https://doi.org/10.1103/PhysRevLett.124.231102} {\bibfield  {journal}
  {\bibinfo  {journal} {Phys. Rev. Lett.}\ }\textbf {\bibinfo {volume} {124}},\
  \bibinfo {pages} {231102} (\bibinfo {year} {2020})},\ \Eprint
  {https://arxiv.org/abs/1912.08988} {arXiv:1912.08988 [gr-qc]} \BibitemShut
  {NoStop}%
\bibitem [{\citenamefont {Chong}\ \emph {et~al.}(2005)\citenamefont {Chong},
  \citenamefont {Cvetic}, \citenamefont {Lu},\ and\ \citenamefont
  {Pope}}]{Chong:2005hr}%
  \BibitemOpen
  \bibfield  {author} {\bibinfo {author} {\bibfnamefont {Z.~W.}\ \bibnamefont
  {Chong}}, \bibinfo {author} {\bibfnamefont {M.}~\bibnamefont {Cvetic}},
  \bibinfo {author} {\bibfnamefont {H.}~\bibnamefont {Lu}},\ and\ \bibinfo
  {author} {\bibfnamefont {C.~N.}\ \bibnamefont {Pope}},\ }\href
  {https://doi.org/10.1103/PhysRevLett.95.161301} {\bibfield  {journal}
  {\bibinfo  {journal} {Phys. Rev. Lett.}\ }\textbf {\bibinfo {volume} {95}},\
  \bibinfo {pages} {161301} (\bibinfo {year} {2005})},\ \Eprint
  {https://arxiv.org/abs/hep-th/0506029} {arXiv:hep-th/0506029} \BibitemShut
  {NoStop}%
\bibitem [{\citenamefont {Cvetic}\ \emph {et~al.}(2022)\citenamefont {Cvetic},
  \citenamefont {Pope}, \citenamefont {Whiting},\ and\ \citenamefont
  {Zhang}}]{Cvetic:2022mal}%
  \BibitemOpen
  \bibfield  {author} {\bibinfo {author} {\bibfnamefont {M.}~\bibnamefont
  {Cvetic}}, \bibinfo {author} {\bibfnamefont {C.~N.}\ \bibnamefont {Pope}},
  \bibinfo {author} {\bibfnamefont {B.~F.}\ \bibnamefont {Whiting}},\ and\
  \bibinfo {author} {\bibfnamefont {H.}~\bibnamefont {Zhang}},\ }\href
  {https://doi.org/10.1103/PhysRevD.106.104026} {\bibfield  {journal} {\bibinfo
   {journal} {Phys. Rev. D}\ }\textbf {\bibinfo {volume} {106}},\ \bibinfo
  {pages} {104026} (\bibinfo {year} {2022})},\ \Eprint
  {https://arxiv.org/abs/2207.01653} {arXiv:2207.01653 [gr-qc]} \BibitemShut
  {NoStop}%
\end{thebibliography}%

\end{document}